\documentclass[referee]{raa}            

\usepackage{graphicx,times}             
\usepackage{natbib}
\usepackage{amssymb,amsmath}
\bibpunct{(}{)}{;}{a}{}{,}

\usepackage[a4paper=true,pagebackref=true]{hyperref}
\hypersetup{colorlinks = true, linkcolor = green, anchorcolor = red, citecolor = blue, filecolor = red, pagecolor = red, urlcolor = red}

\begin{document}

   \title{Energy correction based on fluorescence attenuation of DAMPE
}

   \volnopage{Vol.0 (20xx) No.0, 000--000}      
   \setcounter{page}{1}          

   \author{Libo Wu
         \inst{1,2}
   \and Yunlong Zhang
         \inst{1,2}
   \and Zhiyong Zhang
         \inst{1,2}
   \and Yifeng Wei
         \inst{1,2}
   \and Sicheng Wen
         \inst{1,2}
   \and Haoting Dai
         \inst{1,2}
   \and Chengming Liu
         \inst{1,2}
   \and Xiaolian Wang
         \inst{1,2}
   \and Zizong Xu
         \inst{1,2}
   \and Guangshun Huang
         \inst{1,2}
   }
 

   \institute{State Key Laboratory of Particle Detection and Electronics, University of Science and Technology of China, Hefei, Anhui 230026, China; {\it zhzhy@ustc.edu.cn, weiyf@ustc.edu.cn}\\
        \and
             Department of Modern Physics, University of Science and Technology of China, Hefei, Anhui 230026, China\\
\vs\no
   {\small Received~~20xx month day; accepted~~20xx~~month day}}

\abstract{ The major scientific goals of DArk Matter Particle Explorer (DAMPE) are to study cosmic-ray electrons (including positrons) and gamma rays from 5 GeV to 10 TeV and nuclei from Z = 1 to 26 up to 100 TeV. The deposited energy measured by the Bismuth Germanate Oxide (BGO) calorimeter of DAMPE is affected by fluorescence attenuation in BGO crystals that are 600 mm long. In this work, an in-orbit attenuation calibration method is reported, and energy correction of the sensitive detector unit of the BGO calorimeter is also presented.
\keywords{DAMPE, BGO calorimeter, Fluorescence attenuation, Dark Matter}
}

   \authorrunning{L.- B. Wu et al.}            
   \titlerunning{Energy correction based on fluorescence attenuation of DAMPE}  

   \maketitle

%
%
\section{Introduction}           
\label{sect:intro}
The DArk Matter Particle Explorer (DAMPE) (\citealt{Chang+etal+2017,Ambrosi+etal+2017,Ambrosi+etal+2019,Yuan+Feng+2018}), a satellite-borne experiment funded by the Chinese Academy of Sciences, was launched into a sun-synchronous orbit at an altitude of 500 km in December 2015 from the Jiuquan Satellite Launch Center. The scientific objectives of DAMPE include searching for the signature of dark matter particles (\citealt{Gianfranco+etal+2005,Lu+etal+2014}), understanding the mechanisms of particle acceleration operating in astrophysical sources (\citealt{An+etal+2019}), and studying gamma-ray emission from Galactic and extragalactic sources (\citealt{Duan+etal+2019,Xu+etal+2018}).

\begin{figure}[!htb]
\centering
\includegraphics[width=5in] {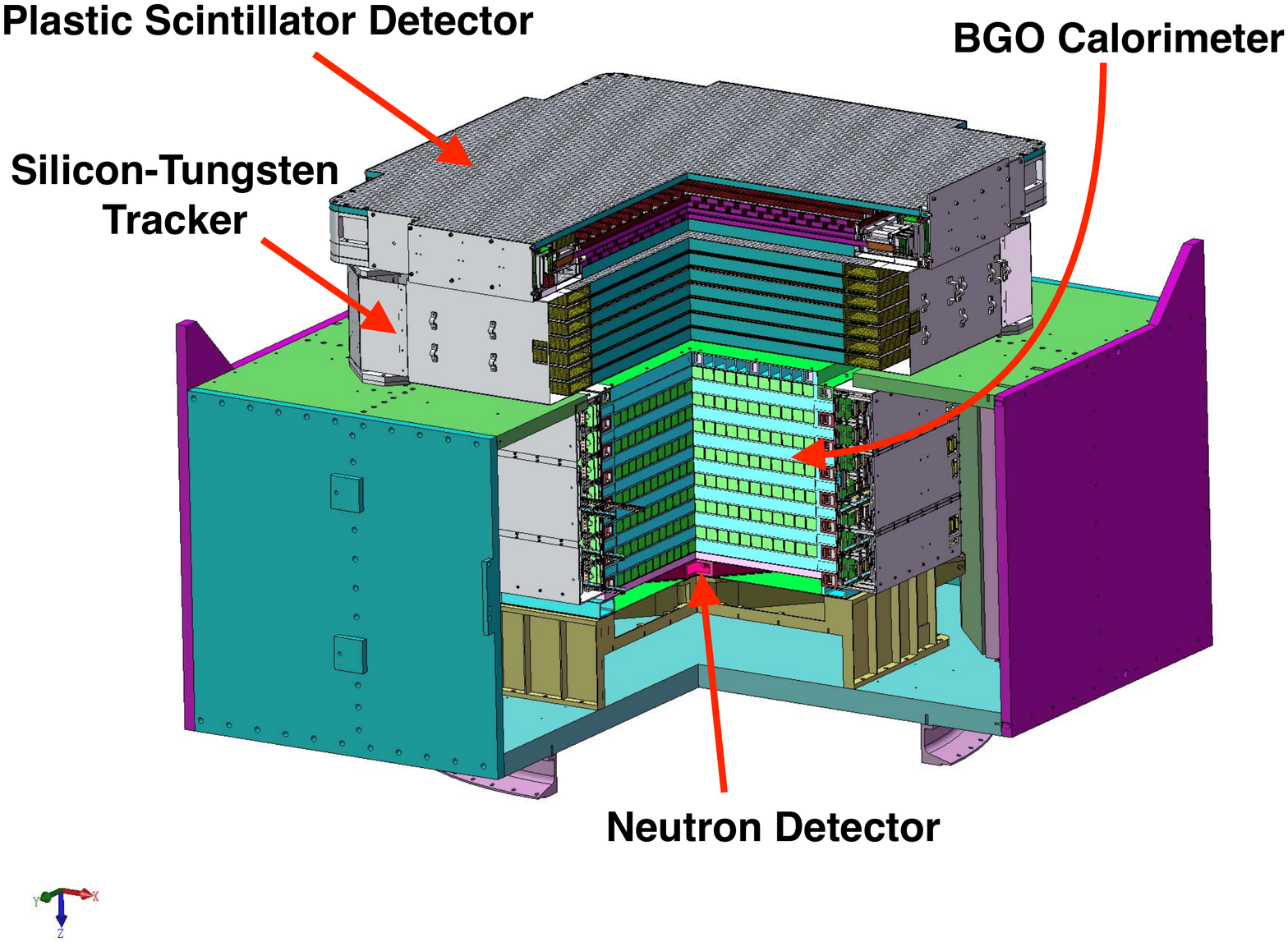}
\caption{Side view of the DAMPE detector.}
\label{fig1}
\end{figure}

DAMPE, as shown in Fig. 1, consists of four sub-detectors. From top to bottom, they are the Plastic Scintillator Detector (PSD) (\citealt{Ding+etal+2019, Ma+etal+2019}), Silicon$-$Tungsten tracKer (STK) (\citealt{Azzarello+etal+2016}), Bismuth Germanium Oxide (BGO) calorimeter (\citealt{Wu+etal+2018,Zhang+etal+2015,Zhang+etal+2019}), and NeUtron Detector (NUD) (\citealt{He+etal+2016}). The PSD, which is composed of two orthogonal layers of plastic scintillator strips with dimensions of 884 $\times$ 28 $\times$ 10 mm, is designed to measure the charge of incoming nuclei via the Z$^2$ dependence of the specific ionization loss in a double layer up to Z = 26, and it  aids in the discrimination between charged particles and photons. The STK is composed of six orthogonal layers of position-sensitive silicon microstrip detectors. Three layers of tungsten are inserted inside layers 2, 3, and 4 to convert gamma rays in electron$-$positron pairs. The STK enables reconstruction of the trajectory and charge of an event. The BGO calorimeter contains 14 layers with 31 radiation lengths and $\sim$1.6 nuclear interaction lengths in total (\citealt{Wei+etal+2016}). Each layer has 22 BGO bars, which are arranged horizontally. The BGO bars of neighboring layers are arranged in an orthogonal way to measure the energy deposition and profile of hadron and electromagnetic showers developed in the BGO calorimeter. Moreover, the BGO provides the trigger for the whole DAMPE system. The NUD is composed of four boron-loaded plastic scintillators, each with a set of photomultiplier tubes (PMTs) and related electronics. It provides additional electron$-$hadron discrimination, which is important for energies above TeV.

Because of the thick BGO calorimeter (\citealt{Wu+etal+2018}), the energy range of DAMPE can cover  from 5 GeV to 10 TeV for electrons and gamma rays and from tens of GeV to hundreds of TeV for cosmic protons and heavier nuclei. The BGO calorimeter is composed of 308 BGO crystal bars with dimensions 25 $\times$ 25 $\times$ 600 mm (produced by the Shanghai Institute of Ceramics), which is the longest BGO crystal available at the time of development. The deposited energy in a BGO bar is independently measured by means of the fluorescent quantities collected by the PMTs at the two ends. The fluorescent photons transmitted along the BGO bar are unavoidably attenuated, so energy correction for this fluorescent attenuation must be taken into account.


\section{Attenuation length calibration}
\label{sect:Att}

\begin{figure}[!htb]
\centering
\includegraphics[width=4in] {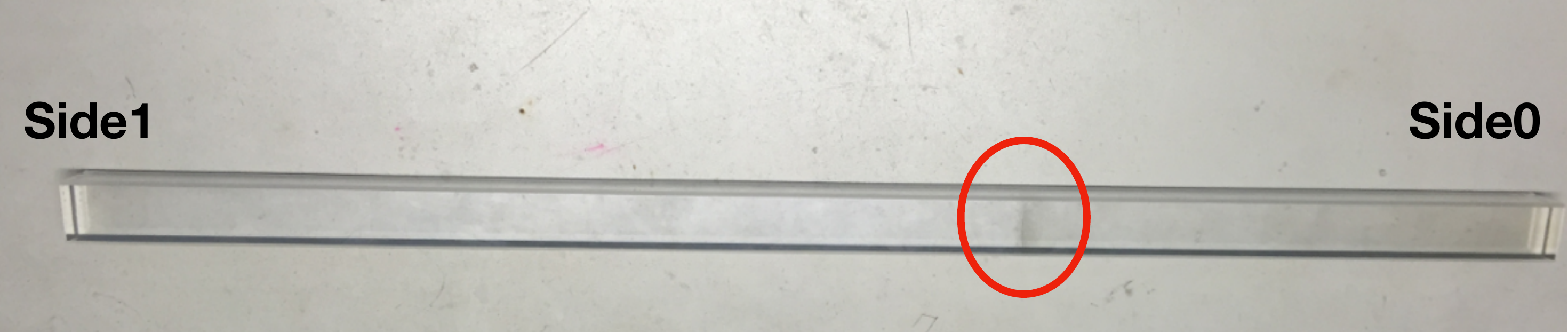}
\caption{BGO crystal with dimension of 25 $\times$ 25 $\times$ 600 mm.}
\label{fig2}
\end{figure}

 Figure 2 shows a 600-mm-long BGO crystal bar (\citealt{Ji+etal+2014}). PMTs are connected to the BGO crystal at both ends (Side0 and Side1) (\citealt{Zhang+etal+2015,Zhang+etal+2012,Feng+etal+2015}) to collected fluorescent lights, which is exponentially attenuated with the distance between the hit position and the PMT. Limited by crystal growth technology at the time of development, there have to be a black shadow called a cutoff point marked by a red cycle in Fig. 2; When scintillating light pass through this region, the absorption and reflection efficiency of the spot to them is very different from other regions. This will cause the energy ratio of reconstruction at both ends to jump here, as shown in Fig. 4.
 In order to eliminate this effect, this paper developed a method to calibrate the attenuation lengths of the crystal at both sides of the spot respectively, and apply them to energy correction.
  
\subsection{Method of calibration}
By considering the energy resolution and the flux of incident particles, all cosmic-ray events are used to calibrate the scintillation attenuation length. The following event selection criteria should be applied:
\begin{itemize}
\item Energy requirement: The total energy deposited in the BGO calorimeter should be $>$ 10 GeV.
\item Track requirement: The BGO track should match the STK track to get an accurate position measurement of the track
\item BarID match requirement: The BGO crystal hit by the track of each layer must have maximum energy deposition in the layer.
\item Acceptance requirement: The events should be in the range of the acceptance angle of the DAMPE detector.
\end{itemize}

\begin{figure}[!htb]
	\begin{minipage}[t]{0.5\linewidth}\centering
	\includegraphics[width=2.7in] {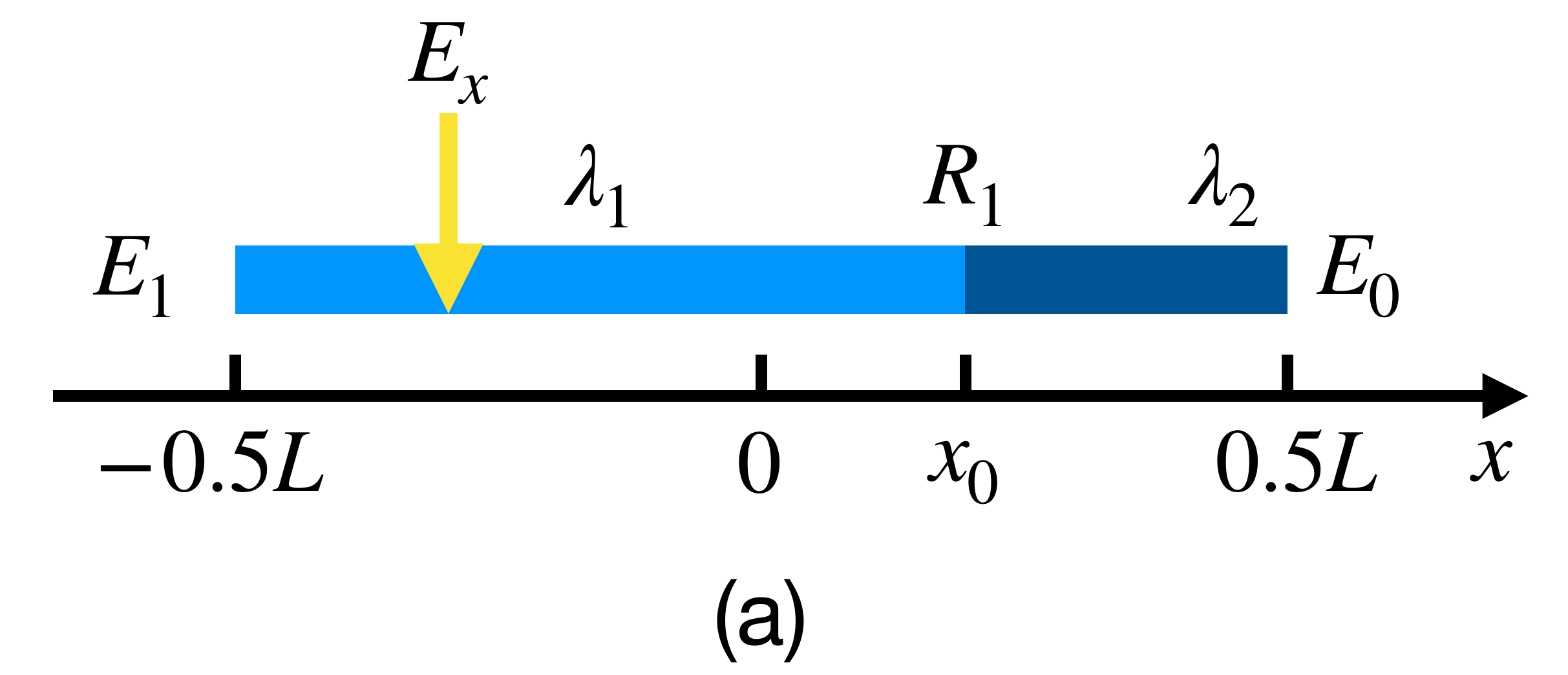}
	\end{minipage}
	\begin{minipage}[t]{0.5\linewidth}\centering
	\includegraphics[width=2.7in] {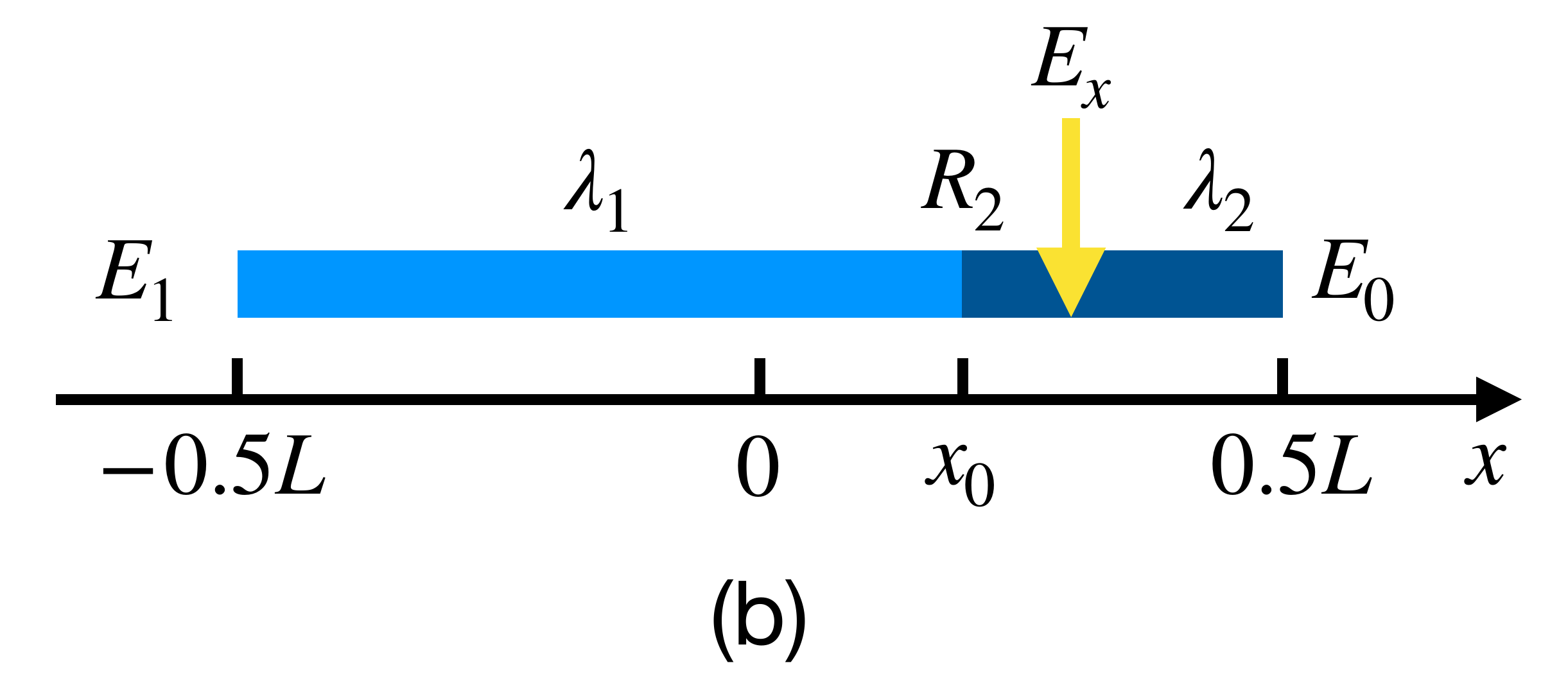}	
	\end{minipage}
	\caption{Schematic plots of a BGO bar with its parameters of interest, showing (a) a hit position on the left side and (b) a hit position on the right side.}
\end{figure}

Figure 3 illustrates the true deposited energy E$_x$ at hit position x, energies E$_0$, E$_1$ measured from the side 0 and side 1, independently, and the cutoff point x$_0$. The origin of the x axis is set at the middle of the BGO bar of total length L = 600 mm. $\lambda_1$ and $\lambda_2$ are the fluorescent attenuation lengths of the two parts ($-$0.5L to x$_0$) and (x$_0$ to 0.5L), respectively. 
There are two cases: Either the hit position is to the left (Fig. 3(a)) of the cutoff point or it is to the right (Fig. 3(b)); the calibration methods are similar. For the case of Fig. 3(a), following the law of fluorescent attenuation, the measured energies $E_0$ and $E_1$, which are proportional to the fluorescent quantities collected by the PMTs of side 0 and 1, can be expressed as:

\begin{equation}
E_{0}=E_{x}  \times e^{-\frac{\left(x_{0}-x\right)}{\lambda_{1}}} \times e^{-\frac{\left(0.5 L-x_{0}\right)}{\lambda_{2}}} \times R_{1}
\end{equation}
\begin{equation}
{E_{1}=E_{x} \times e^{-\frac{x+0.5 L}{\lambda_{1}}}}
\end{equation}

where E$_{x}$ is the true deposited energy at position x (from $-$300 to 300 mm) in the BGO bar and $R_1$ in formula (1) is the transparent coefficient when fluorescent photons pass through the cut point $x_0$ from the left to right.

The attenuation length of the left part (x $< x_0$) can be calibrated with the following function derived from functions (1) and (2):
\begin{equation}
\ln \left(\frac{E_{0}}{E_{1}}\right)=\frac{2 x}{\lambda_{1}}+\left[\frac{0.5 L-x_{0}}{\lambda_{1}}+\frac{x_{0}-0.5 L}{\lambda_{2}}+\ln R_{1}\right]
\end{equation}
Similarly, for the cases when particles are incident on the right ($x > x_0$), the attenuation length $\lambda_2$ can be calibrated by using formula (4), where $R_2$ is the transparent coefficient when fluorescent photons pass through the cutoff point $x_0$ from right to left.
\begin{equation}
\ln \left(\frac{E_{0}}{E_{1}}\right)=\frac{2 x}{\lambda_{2}}+\left[\frac{0.5 L+x_{0}}{\lambda_{1}}-\frac{0.5 L+x_{0}}{\lambda_{2}}+\ln \frac{1}{R_{2}}\right]
\end{equation}

\begin{figure}[!htb]
\centering
\includegraphics[width=3.6in] {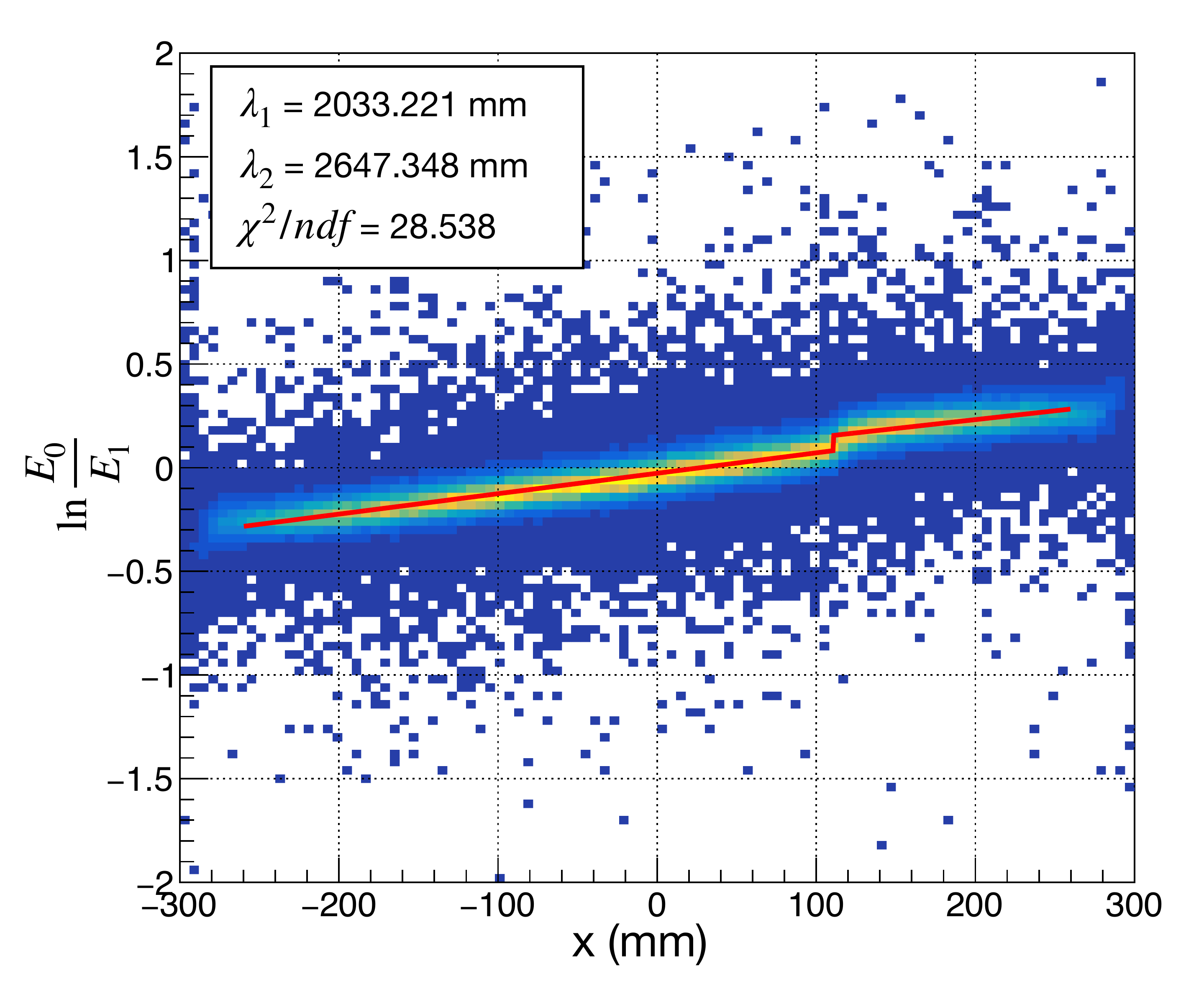}
\caption{Scatterplot of $\ln \frac{E_0}{E_1}$ versus hit position x. The fit range is from $-$260 mm to 260 mm.}
\label{fig2}
\end{figure}
Taking an example of the attenuation length calibration of a BGO bar (No. 10, layer 3), the hit position x and the measured energies $E_0$ and $E_1$ are, event by event, reconstructed according the event track fitting and ADC counts readout from electronics of the both sides. Fig.4 shows a typical scatterplot of $\ln{\frac{E_0}{E_1}}$ versus x of the BGO bar at No.10 of layer 3.


\subsection{Calibration results}
Fitting the scatterplot of Fig. 4 with the formulas (3) and (4) gives fluorescent attenuation lengths of the BGO bar of $\lambda_1$ = 2033 mm and $\lambda_2$ = 2647 mm. The parameters $R_1$ and $R_2$ can be calculated from the intercept of the fitting line. The same calibration program was applied to all 308 piecse of the BGO bars. Figure 5 and 6 show the distributions of $\lambda_1$ and $\lambda_2$ and $R_1$ and $R_2$, respectively, for the BGO calorimeter.

\begin{figure}[!htb]
	\begin{minipage}[t]{0.5\linewidth}\centering
	\includegraphics[width=2.5in] {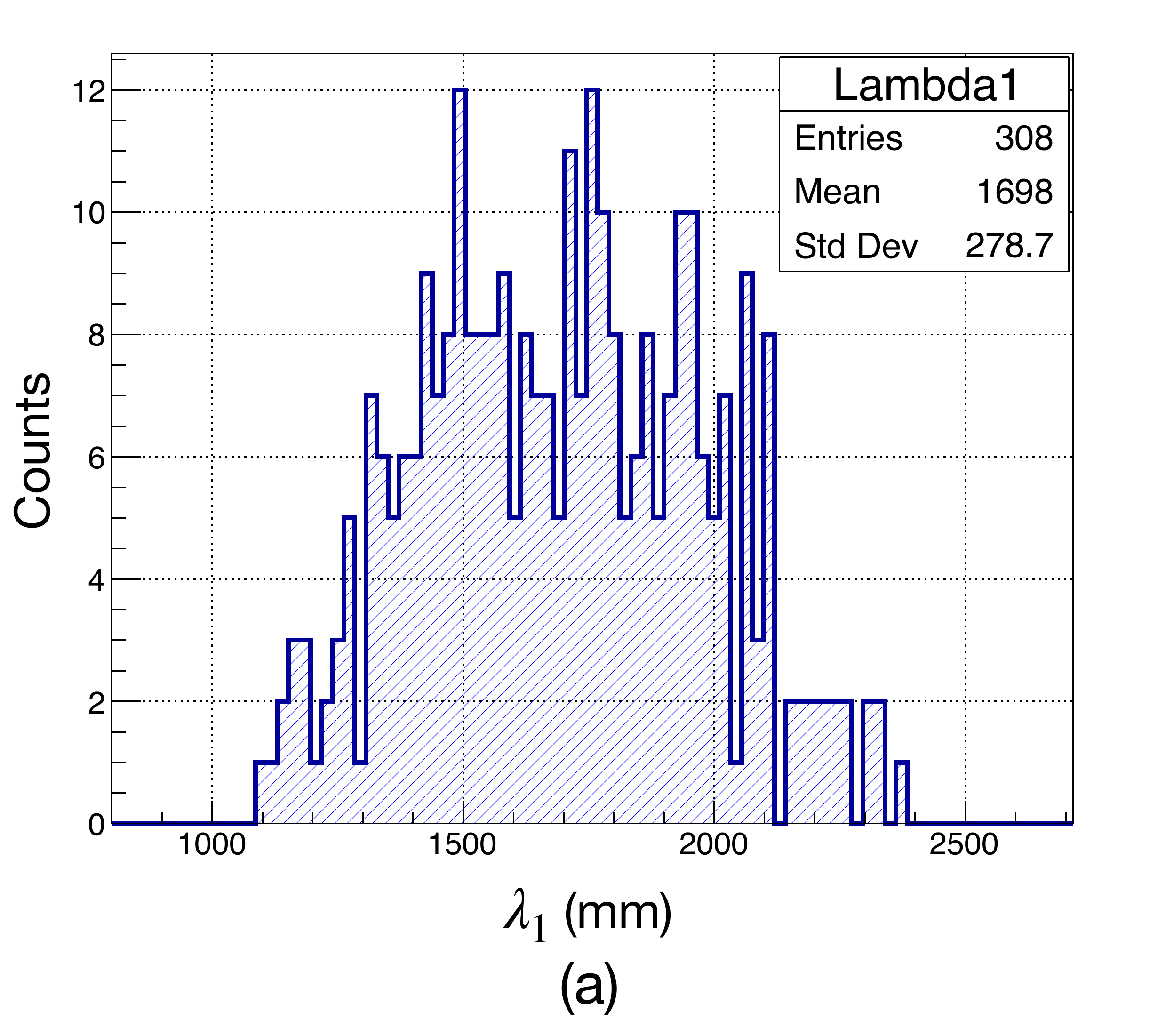}
	\end{minipage}
	\begin{minipage}[t]{0.5\linewidth}\centering
	\includegraphics[width=2.5in] {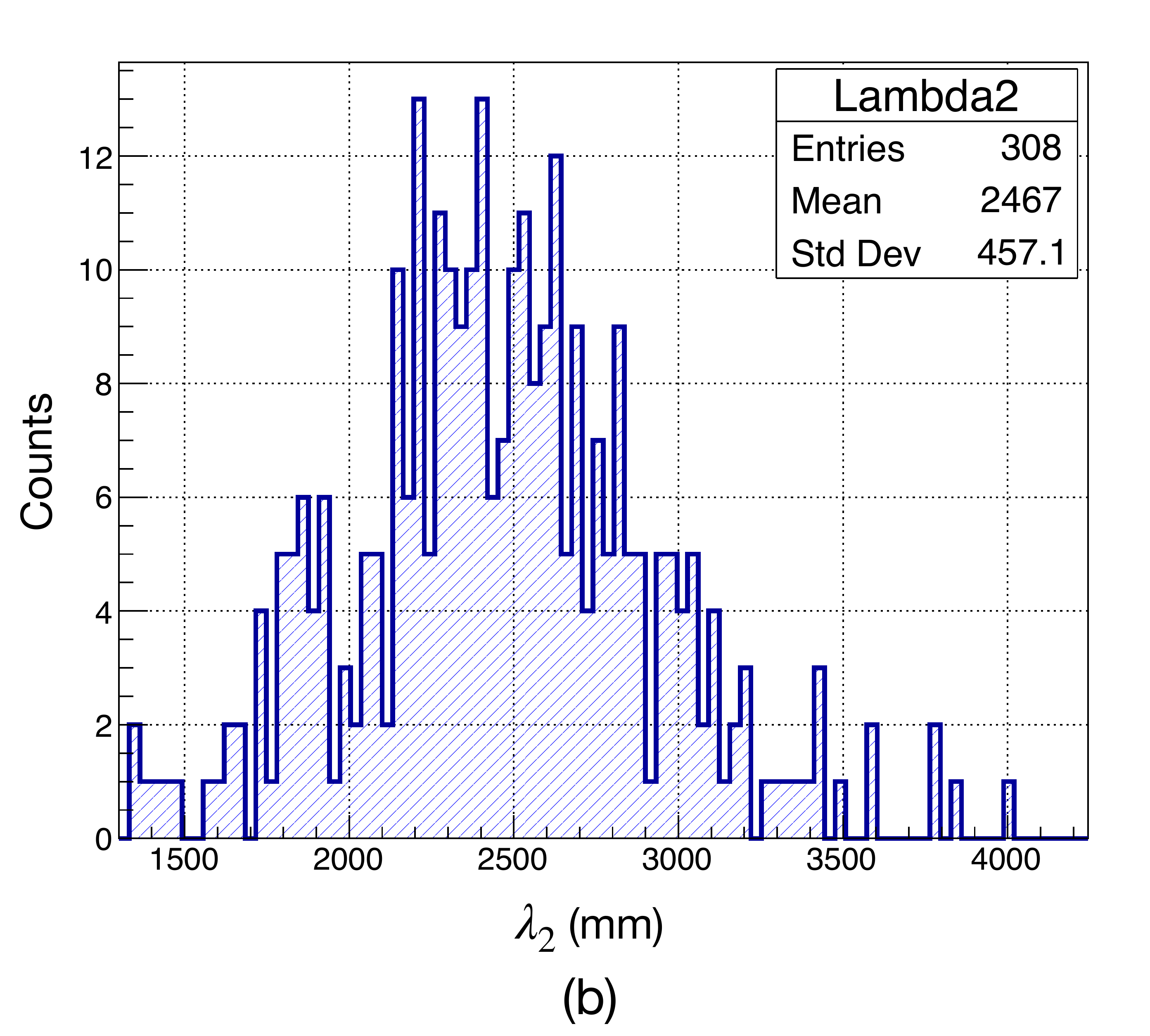}	
	\end{minipage}
	\caption{Distributions of attenuation lengths of all BGO crystals: (a) $\lambda_1$; (b).}
\end{figure}
\begin{figure}[!htb]
	\begin{minipage}[t]{0.5\linewidth}\centering
	\includegraphics[width=2.5in] {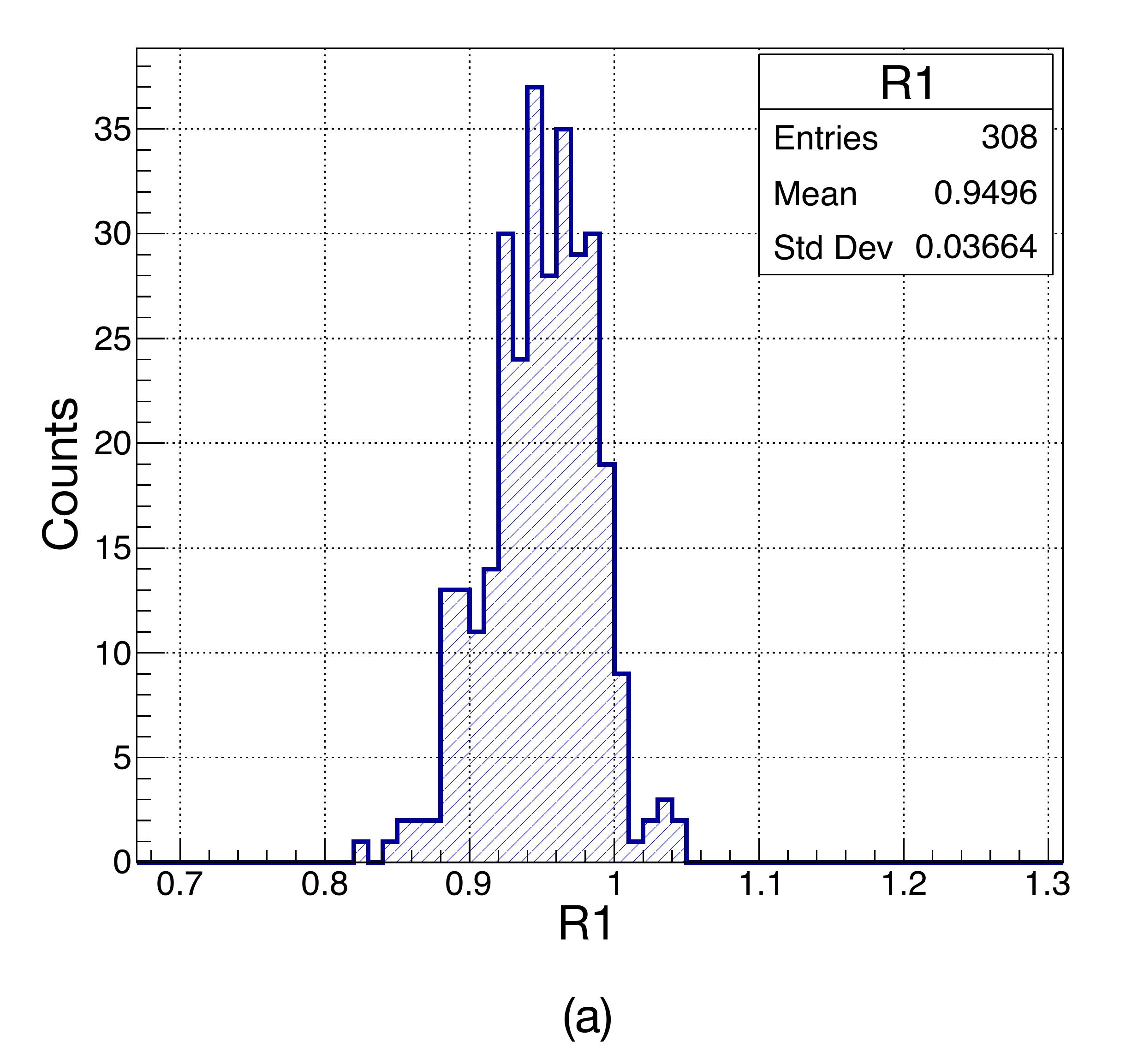}
	\end{minipage}
	\begin{minipage}[t]{0.5\linewidth}\centering
	\includegraphics[width=2.5in] {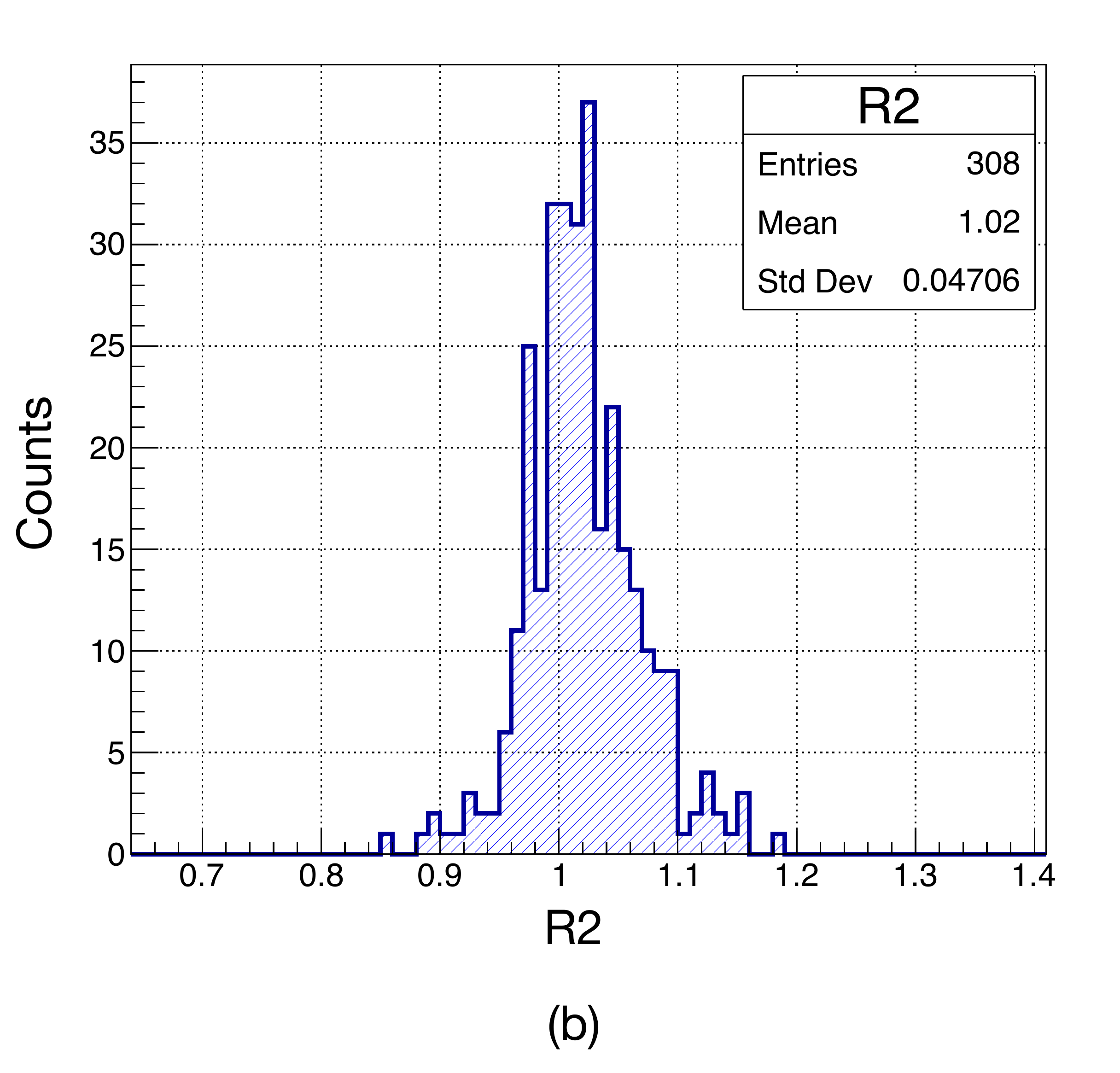}	
	\end{minipage}
	\caption{Distribution ratios of the variation of scintillation intensity. (a) $R_1$; (b) $R_2$.}
\end{figure}
A database of the calibration parameters of the 308 pieces of the BGO bar ($\lambda_1$, $\lambda_2$ and $R_1$, $R_2$, $x_0$) has been set up for the program of the following energy correction.

\section{Energy correction based on attenuation length}
\label{sect:ECorrection}
In the experiment, the measured energies $E_0$ and $E_1$ are different from the true deposited energy at the hit position x because of the fluorescent attenuation, as shown in formulas (1) and (2). Therefore, it is necessary to perform an energy correction to account for fluorescence attenuation for each BGO crystal.

\subsection{Method of energy correction}
 By using the database, the true deposited energy at hit position x can be, event by event, corrected for each BGO bar by means of the following formulas:

 \begin{equation}
E_{x_0}=E_{0} \times e^{\frac{x_{0}-x}{\lambda_{1}}} \times e^{\frac{0.5 L-x_{0}}{\lambda_{2}}} \times \frac{1}{R_{1}}
\end{equation}
 \begin{equation}
 E_{x_1}=E_{1} \times e^{\frac{x+0.5 L}{\lambda_{1}}}
 \end{equation}
  \begin{equation}
E_{x_0}=E_{0} \times e^{\frac{0.5 L-x}{\lambda_{2}}}
 \end{equation}
 \begin{equation}
E_{x_1}=E_{1} \times e^{\frac{x-x_{0}}{\lambda_{2}}} \times e^{\frac{x_{0}+0.5 L}{\lambda_{1}}} \times \frac{1}{R_{2}}
 \end{equation}
 
where formulas (5) and (6) are energy correction for cases when particles hit the left part (x $< x_0$) of the BGO crystal, while (7) and (8) are for cases when particles hit the right part (x $\geq x_0$). $E_{x_0}$ and $E_{x_1}$ are the energies at side 0 and  side 1 after the attenuation correction, respectively.  Moreover, the hit position reconstructed by track of each layer is used to correct the energy of each BGO bars of the corresponding layer.

\begin{figure}[!htb]
\centering
\includegraphics[width=3.5in] {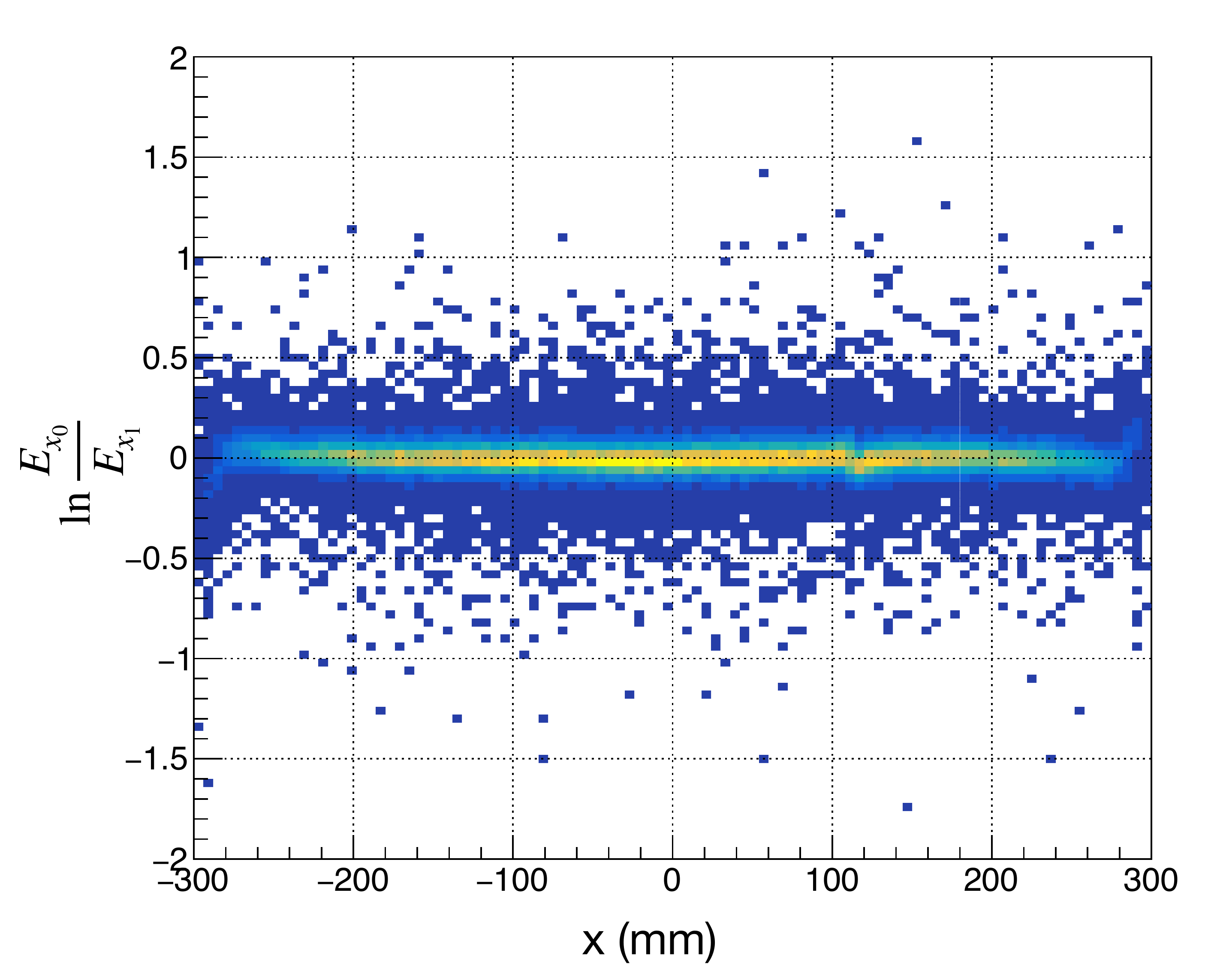}
\caption{Scatterplot of $\ln \frac{E_{x_0}}{E_{x_1}}$ versus hit position of x after energy correction.}
\label{fig2}
\end{figure}

\subsection{Energy correction results}
The correlation between $\ln \frac{E_0}{E_1}$ and hit position (x) of the BGO bar (No. 10, layer 3) after correction is shown in Fig. 7. Note that the distribution is flatter than that in Fig. 4, without a jump at $x_0$. Figure 8 shows a comparison of the relationship of the energies at the two sides before (Fig. 8(a)) and after (Fig. 8(b)) correction. The plot in Fig. 8(b) demonstrates that, after energy correction, $E_{x_0}$ and $E_{x_1}$ measured at the two sides are getting closer to each other and become two independent measures of the true deposited energy at hit position x. Therefore, the correction program will increase the energy measurement accuracy and redundancy.

\begin{figure}[!htb]
	\begin{minipage}[t]{0.5\linewidth}\centering
	\includegraphics[width=3in] {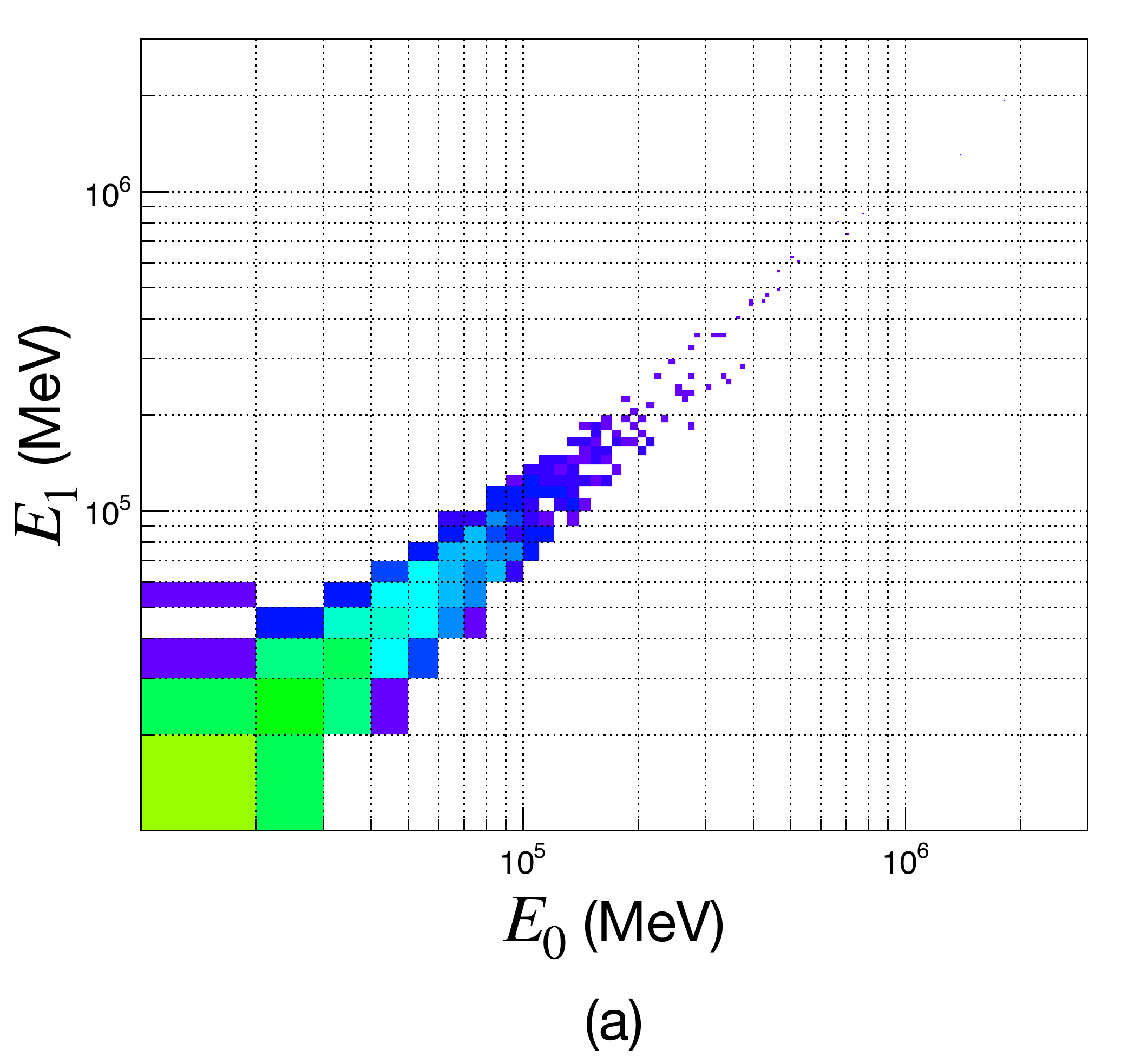}
	\end{minipage}
	\begin{minipage}[t]{0.5\linewidth}\centering
	\includegraphics[width=3in] {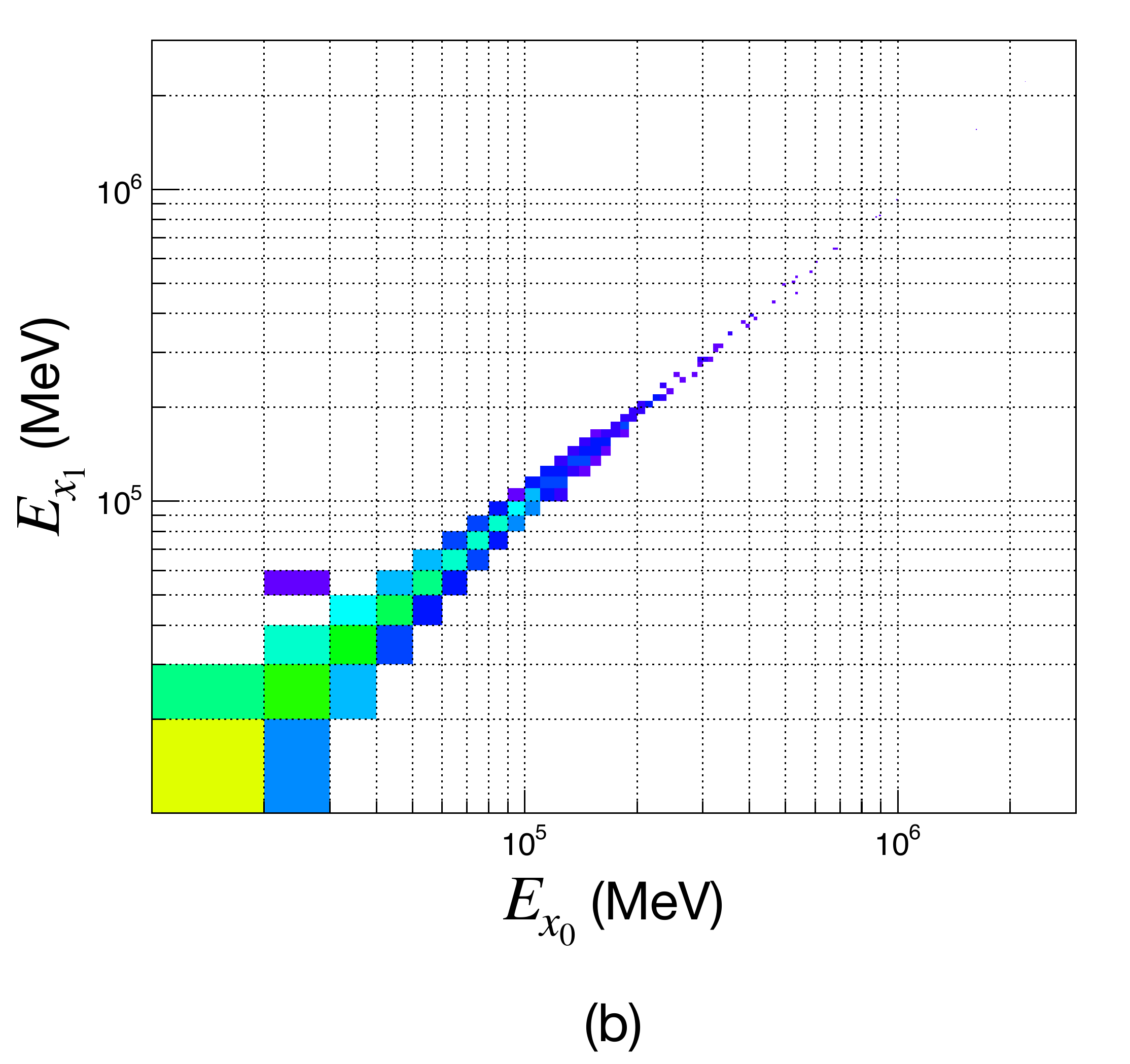}	
	\end{minipage}
	\caption{Comparison of the relationships of the energies at the two sides (a) before correction and (b) after correction.}
\end{figure}

\begin{figure}[!htb]
	\begin{minipage}[t]{0.5\linewidth}\centering
	\includegraphics[width=3in] {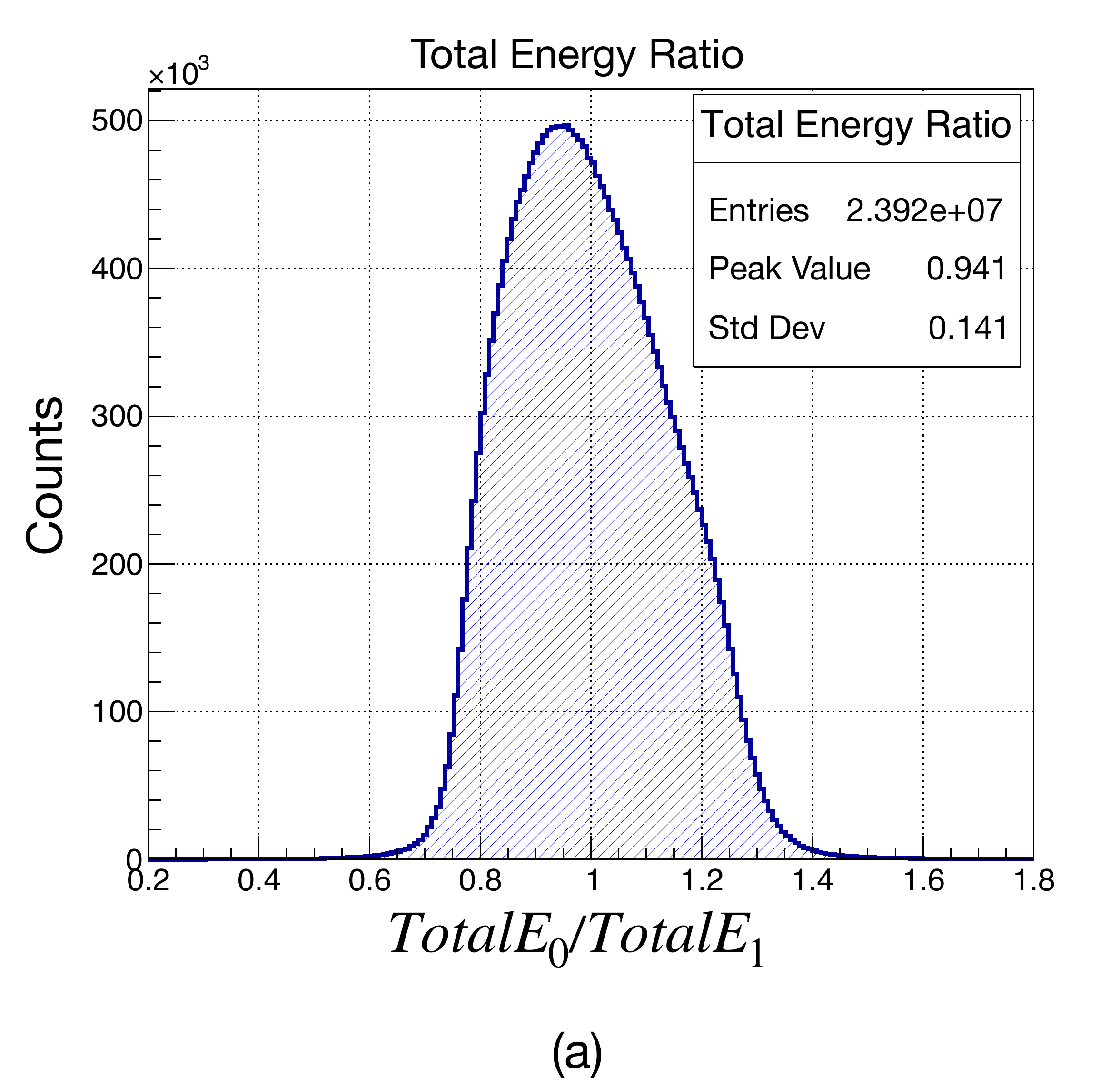}
	\end{minipage}
	\begin{minipage}[t]{0.5\linewidth}\centering
	\includegraphics[width=3in] {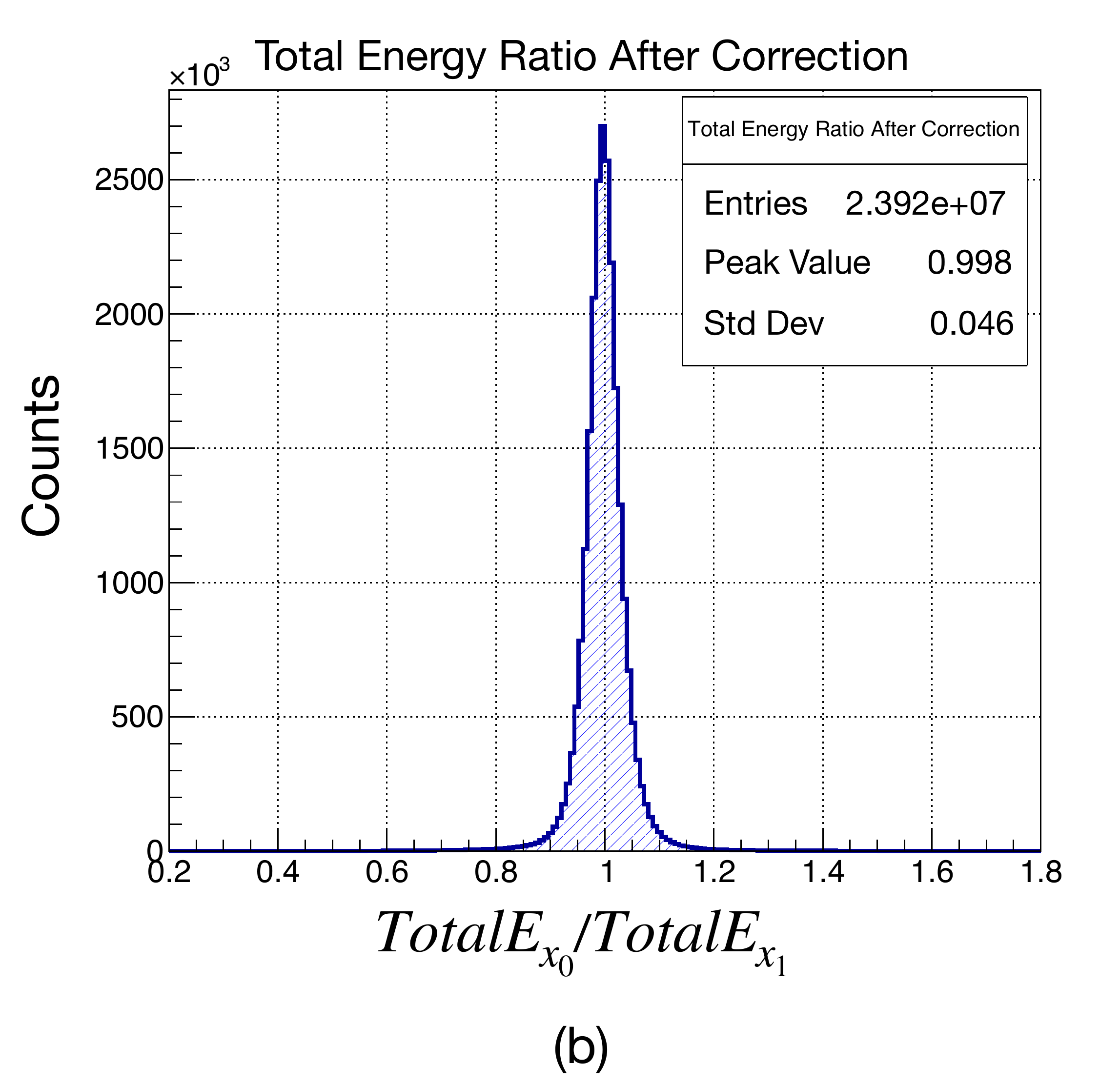}	
	\end{minipage}
	\caption{Ratios of the total deposited energies reconstructed from the side 0 and side 1 of all BGO crystals: (a) before correction; (b) after correction.}
\end{figure}

The total deposited energy can be obtained by using the correction method on all BGO crystals, and the distributions of the ratio of total deposited energy at the two sides before and after correction are shown in Figs. 9(a) and 9(b), respectively. The peak value is 0.998 after correction, which is more closer to 1 than the result before correction, and the standard deviation after correction is 0.046, which is much less than the result of 0.141 before correction.

\section{Conclusion}
\label{sect:conclusion}
There really is fluorescence attenuation in BGO crystal, which will influence the energy reconstructed at both sides. An in-orbit method to calibrate attenuation length based on the unique BGO crystals of DAMPE has been studied, and a database of calibration parameters has been set up. Upon completion of the energy correction for fluorescence attenuation, the consistency between energy measurements at the two sides was raised from 0.941 to 0.998, $E_{x_0}$ and $E_{x_1}$ became two independent measurements of the true deposited energy $E_x$, and the energy resolution and redundancy capability of the BGO calorimeter were significantly improved.

\begin{acknowledgements}
This work is supported by the Project supported by the Joint Funds of the National Natural Science Foundation of China (No.  U1738135, U1738208 and U1738139), the National Natural Science Foundation of China (No. 11673021 and No. 11705197) and the National Key Research and Development Program of China (2016YFA0400200 and 2016YFA0400202).
\end{acknowledgements}

\bibliographystyle{raa}
\bibliography{bibtex}










\end{document}